# WAND: A 128-channel, closed-loop, wireless artifact-free neuromodulation device

Andy Zhou[1]*, Samantha R. Santacruz[1,2]*, Benjamin C. Johnson[1,3]*, George Alexandrov[1], Ali Moin[1], Fred L. Burghardt[1], Jan M. Rabaey[1t], Jose M. Carmena[1,2t], Rikky Muller[1,3t]

[1] Department of Electrical Engineering and Computer Sciences, University of California, Berkeley, Berkeley, CA USA 94720

[2] Helen Wills Neuroscience Institute, University of California, Berkeley, Berkeley, CA USA 94720

[3] Cortera Neurotechnologies, Inc., Berkeley, CA USA 94704

Correspondence should be addressed to R.M. (rikky@berkeley.edu)

## Abstract

Closed-loop neuromodulation systems aim to treat a variety of neurological conditions by dynamically delivering and adjusting therapeutic electrical stimulation in response to a patient's neural state, recorded in real-time. Existing systems are limited by low channel counts, lack of algorithmic flexibility, and distortion of recorded signals from large, persistent stimulation artifacts. Here, we describe a device that enables new research applications requiring high-throughput data streaming, low-latency biosignal processing, and truly simultaneous sensing and stimulation. The Wireless Artifact-free Neuromodulation Device (WAND) is a miniaturized, wireless neural interface capable of recording and stimulating on 128 channels with on-board processing to fully cancel stimulation artifacts, detect neural biomarkers, and automatically adjust stimulation parameters in a closed-loop fashion. It combines custom application specific integrated circuits (ASICs), an on-board FPGA, and a low-power bidirectional radio. We validate wireless, long-term recordings of local field potentials (LFP) and real-time cancellation of stimulation artifacts in a behaving nonhuman primate (NHP). We use WAND to demonstrate a closed-loop stimulation paradigm to disrupt movement preparatory activity during a delayed-reach task in a NHP *in vivo*. This wireless device, leveraging custom ASICs for both neural recording and electrical stimulation modalities, makes possible a neural interface platform technology to significantly advance both neuroscientific discovery and preclinical investigations of stimulation-based therapeutic interventions.

Closed-loop neuromodulation improves open-loop therapeutic electrical stimulation by providing adaptive, on-demand therapy, reducing side effects, and extending battery life in wireless devices[1,2]. Closing the loop requires low-latency extraction and accurate estimation of neural biomarkers[3–5] from recorded signals to automatically adjust when and how to administer stimulation as feedback to the brain. Recent studies have shown responsive stimulation to be a viable option for treating epilepsy[2,6], and there is evidence that closed-loop strategies could improve deep brain stimulation (DBS) for treating Parkinson's disease and other motor disorders[7,8]. However, there is presently no commercial device allowing closed-loop stimulation for DBS in patients with movement disorders, and strategies for implementing such stimulation are still under investigation. In fact, most attempts to close the loop for DBS treatments has been done only for short duration using non-fully implantable systems[5,9–12]. To enable advanced research in closed-loop neuromodulation, there is a need for a flexible research platform for testing and implementing these various closed-loop paradigms that is also wireless, compact, robust, and safe.

Designing such a device requires unification of multi-channel recording, biomarker detection, and microstimulation technologies into a single unit with careful consideration of their interactions. Wireless, multi-channel recording-only devices capture activity from wide neuronal populations[13,14], but do not have the built-in ability to immediately act on that information and deliver stimulation. Several complete closed-loop devices have been proposed and demonstrated, but are limited by low channel-counts[15–18] and low wireless streaming bandwidth[15–19]. Most recently, a number of fully-integrated and optimized closed-loop neuromodulation SoCs have been presented, but full system functionality has not yet been adequately demonstrated in vivo[20–26]. While future versions may be paired with miniaturized external battery packs and controllers, current systems built around these SoCs require large, stationary devices to deliver power inductively from a close range[20,22,24]. This limits studies to using small, caged animals.

Furthermore, any device for concurrent sensing and stimulation must be able to mitigate or remove stimulation artifacts, the large voltage transients resulting from stimulation that distort recorded signals and obscure neural biomarkers. Signals recorded concurrently with stimulation may contain relevant information for closed-loop algorithms or offline analysis, yet existing devices disregard these affected windows of data, or fail to reduce artifacts to an acceptable level for recovery of many potentially useful biomarker features. Effectively and efficiently cancelling artifacts requires careful co-design of the stimulators and signal acquisition chains. Additionally, computational reprogrammability is needed for application-dependent algorithm design in both artifact cancellation and closed-loop control.

The Wireless Artifact-free Neuromodulation Device (WAND) introduced here is the first to our knowledge to incorporate all key features needed to continuously monitor neural biomarkers in the presence of stimulation artifacts and deliver closed-loop stimulation. WAND combines: (1) two custom, 64-channel neural interface application specific integrated circuits (ASICs) supporting simultaneous low-noise recording and high-current stimulation, specifically designed to minimize stimulation artifact; (2) flexible and reprogrammable back-end processing on a system-on-a-chip field-programmable gate array (SoC FPGA) for cancelling residual artifacts, computing neural biomarkers, running closed-loop algorithms, and controlling stimulation; and (3) a robust, bidirectional wireless link to a graphical user interface (GUI) for device configuration and control, as well as data logging. These features are tightly integrated into a small form factor, low-power device, enabling many proposed closed-loop and responsive neuromodulation applications, as well as offering a platform for developing new ones. To demonstrate the various functions of WAND, we perform a series of *in vivo* experiments that validate long-term, high-fidelity, and wireless multi-channel recording; real-time, complete removal of stimulation artifacts for accurate recovery of neural signals; and on-board biomarker detection for closed-loop control.

## System design motivation

WAND is designed to be a general-purpose tool with immediate applicability in various research environments. Inclusion of a wide feature-set is balanced by limitations in device size and power.

Integrated circuits are required to minimize area and power for a large number of recording and stimulation channels. We designed a custom neuromodulation integrated circuit (NMIC) to deliver stimulation pulses ranging from subthreshold currents (down to 20 µA) to those required by DBS (5 mA), and to record local field potentials (LFPs) with a bandwidth of up to 500 Hz[27]. We chose LFP as the signal of interest for its usefulness in medical applications as an indicator of disease[8,28–31]. There is also evidence that LFP can be used for motor decoding in brain-machine interfaces[32–36], with comparable accuracy and better longevity than spike decoders[36]. This signal is also extremely useful to understanding neural processing and incredibly relevant for a variety of basic neuroscience studies, from investigating how neural oscillations coordinate movement[37–39] to cued transitions between dynamic states in cortico-basal ganglia circuits[40] and working memory[41,42]. Finally, the lower (1 kHz) sampling rate required for LFP recording utilizes lower wireless bandwidth for real-time streaming, allowing the use of low-power, off-the-shelf radios.

While numerous high channel count recording circuits have been designed[43,44], state-of-the-art circuits cannot tolerate, and often exacerbate, the effects of stimulation artifacts. Electrical stimulation generates a large voltage transient (direct artifact), concomitant with stimulation current and charge delivery to neural tissue and nearby electrodes. Direct artifact may be many orders of magnitude larger that the underlying neural signal (mV compared to µV, respectively). This is followed by a long, post-stimulus voltage decay (indirect artifact) determined by the mismatch of stimulation phases and electrode properties. Conventional low-noise, low-power neural amplifiers are sensitive to both direct and indirect artifact, saturating from both[45,46]. They recover slowly from saturation due to long time constants of analog feedback, causing data loss during and many milliseconds after a stimulation pulse. State-of-the-art methods for mitigating the indirect artifact try to prevent saturation of the front-ends. Saturation can be prevented by increasing the amplifier linear input range and tolerance to DC offset[22,27,47], or by subtracting the large amplitude components of the artifact[48,49]. Artifact duration can be reduced by rapidly clearing charge built up on circuit elements from stimulation[26,27,50,51]. We have designed the NMICs with improved stimulation and recording architectures to both prevent large indirect artifacts and minimize their effects on the front-end circuits.

To date, even the best results in front-end artifact mitigation do not demonstrate complete artifact removal, necessitating back-end digital cancellation for residual artifacts. Computationally efficient back-end methods include subtractive methods, where artifact templates are subtracted from the waveform to reveal the underlying signal[49,52,53], and reconstructive methods, where segments of corrupted data are replaced with interpolated values[54–56]. Different techniques may achieve better results than others, depending on the level of mitigation achieved by the front-end amplifiers.

Overall system resiliency to stimulation artifact depends heavily on co-design of the stimulator, recording front-end, and signal processing blocks. In this work, we demonstrate how the specific artifact prevention and mitigation techniques utilized in the NMICs motivate our implementation of back-end linear interpolation in an on-board SoC FPGA to completely remove stimulation artifacts in real time. In particular, the reduced artifact duration allows for a computationally inexpensive but effective back-end cancellation solution at the cost of losing only one or two samples. These innovations allow for online, real-time biomarker computation for closed-loop stimulation.

## Results

### WAND architecture

WAND components and architecture are shown in Fig. 1. For this work, the form factor was designed to fit into the polyetherimide housing for a custom chronically implanted microelectrode array (Gray Matter Research, Bozeman, MT). The device has a board area of 10.13 cm$^2$ and weight of 17.95 g together with a rechargeable 500 mAh Li-ion battery pack, allowing 11.3 hours of continuous, wireless operation (**Fig. 1a**). The main components of WAND are the pair of custom NMICs, an SoC FPGA, a radio SoC, and support circuitry for power regulation and programming (**Fig. 1b, d**).

Each NMIC consists of 64 recording channels and 4 stimulators that can address any of the 64 channels, meaning that stimulation can occur simultaneously on up to 8 channels by leveraging stimulators across the two on-board NMICs. Multi-site stimulation is desirable for implementing specific spatiotemporal patterns of stimulation, with many recent studies performing stimulation on 2 – 8 channels simultaneously[57–62]. Using WAND, the stimulation channels can be dynamically assigned, thus allowing this device to be utilized for multi-site stimulation on up to 8 channels concurrently in a highly flexible manner. Ultimately, stimulation can be delivered using an open-loop paradigm or a closed-loop approach that relies on continuous sensing of onboard computed biomarkers.

Stimulation parameters are rapidly reprogrammable by writing to registers on the NMIC through commands transmitted from a GUI or automatically based on calculations performed on-board. All stimulation settings listed in **Supplementary Fig. 1**, as well as selection of stimulation sites and triggering of pulses, can be set through the same interface. A new setting can be preloaded while stimulating with a previous setting, potentially reducing latency between biomarker state detection and the resulting stimulation update.

Unlike conventional neural interface ICs, the NMICs enable simultaneous low-noise, low-power neural recording of LFP with high-compliance electrical stimulation (**Fig. 1c**). The NMIC prevents large amplitude indirect artifacts by employing stimulators with highly accurate charge balancing[27]. Accurate charge balancing is achieved by reusing the same current source for both phases of a biphasic pulse and a return-to-ground stimulator architecture (**Supplementary Fig. 2**). To address both direct and indirect artifact, the NMIC recording front-ends are designed simultaneously for low noise (1.6 $\mu V_{rms}$ mean channel noise) recording and a large linear input range of 100mV. The input range is over 10 times larger than conventional designs[43,44] and avoids saturation in the presence of a large stimulation artifacts (10's of mV) while still being able to resolve µV level neural signals. This large range of resolvable signals is achieved with a mixed-signal architecture that integrates the ADC into the feedback loop, thereby reducing the required gain and signal swings. The architecture also resets at every sample, enabling memoryless sampling and rapid recovery from stimulation artifacts (**Supplementary Fig. 4**). Therefore, stimulation artifacts do not persist beyond the samples when stimulation is occurring, and a minimal amount of data is lost when using reconstructive back-end cancellation methods such as interpolation.

All 128 channels of neural data from both NMICs are sampled, digitized (15 bits, 1 kS/s), and serialized on chip and transmitted to the on-board FPGA and microcontroller SoC via a custom bidirectional interface implemented in the FPGA fabric[63]. The same interface is used for downlink commands to control NMIC circuitry and update stimulation parameters (**Fig. 1d**). Software running on the included Cortex M3 microprocessor then aggregates neural and other sensor data, cancels stimulation artifact, selects a subset of data to be streamed back to the base station, and runs closed-loop neuromodulation algorithms (**Fig. 1d**). The FPGA fabric and Cortex M3 software are reprogrammable through a serial wire debug interface allowing customization for different applications. A 2.4 GHz Bluetooth Low Energy (BLE) radio SoC allows for robust bidirectional wireless communication up to 2 m from the subject. BLE offers low power telemetry, and customization of the BLE protocol enables data streaming rates close to the 2 Mbps modulation rate. WAND can stream up to 96 uncompressed LFP recording channels in real time to a PC running a custom GUI for system configuration and data visualization.

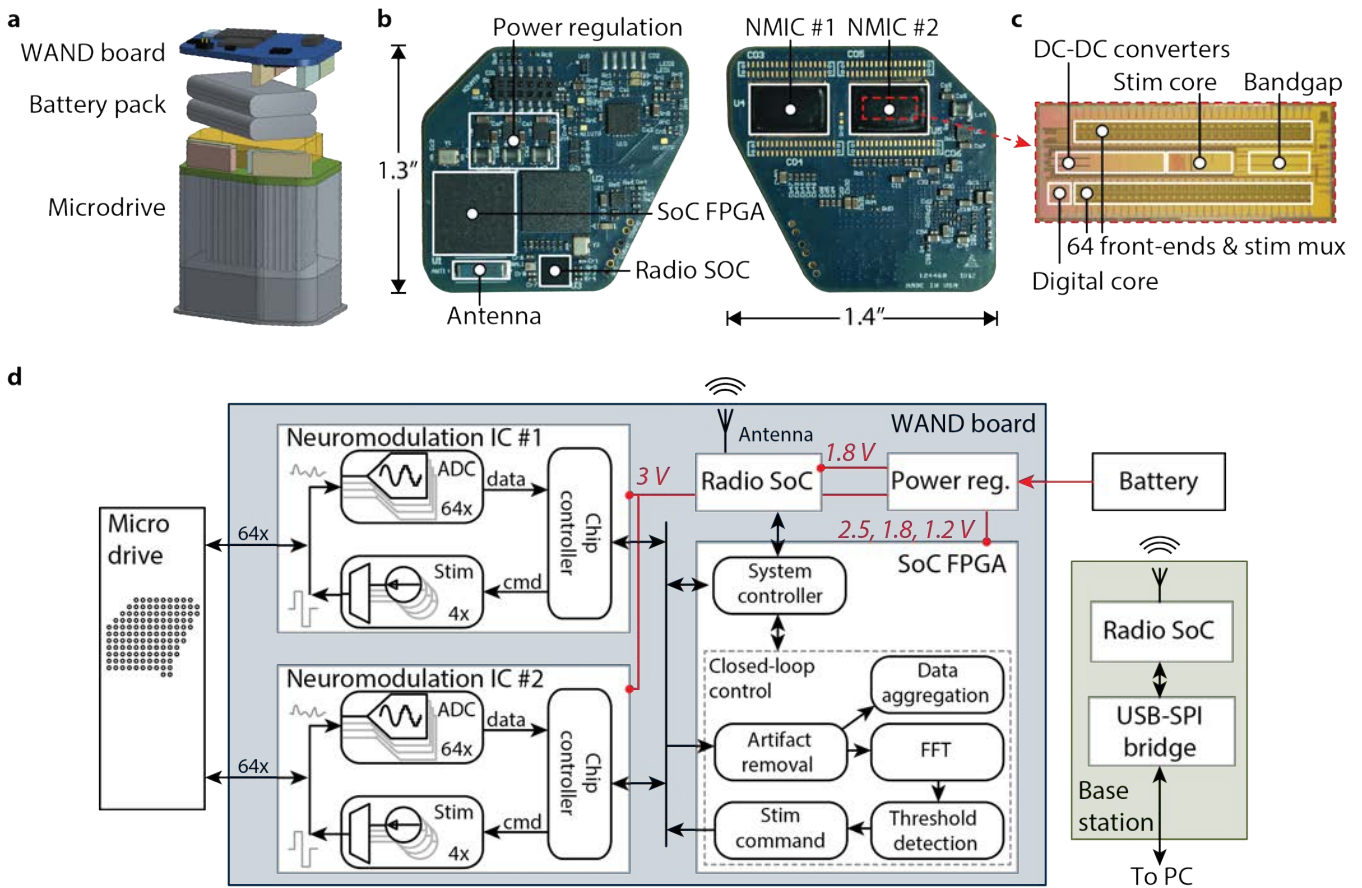

**Figure 1 – WAND system architecture.** (**a**) 3D CAD model of WAND with the primate headstage and battery pack, shown without polyetherimide case. (**b**) Top- and bottom-view photographs of WAND circuit board with relevant subsystems annotated. (**c**) Micrograph of custom neural interface IC (NMIC) with annotated subcircuits. (**d**) Functional diagram of the WAND system showing data and power connections on the main device board, and connections to the microdrive electrode array, battery, and a wireless base station.

### High-fidelity multi-channel wireless recording

To evaluate the quality of recordings made using WAND, we recorded 96 channels of LFP activity from a NHP using a chronically implanted microdrive electrode array with access to both cortical and subcortical nuclei (**Fig. 2a**). We compared WAND recordings with sequentially recorded neural data from a wired, state-of-the-art, commercial neurophysiology system (Tucker-Davis Technologies, Alachua, FL). Respective recordings from each system have qualitatively similar signal properties, as assessed by computing the power spectral densities (PSD) of the recorded data (**Fig. 2b, c**). The WAND recordings exhibit lower 60 Hz interference due to the lack of long interface cables and better isolated recording references.

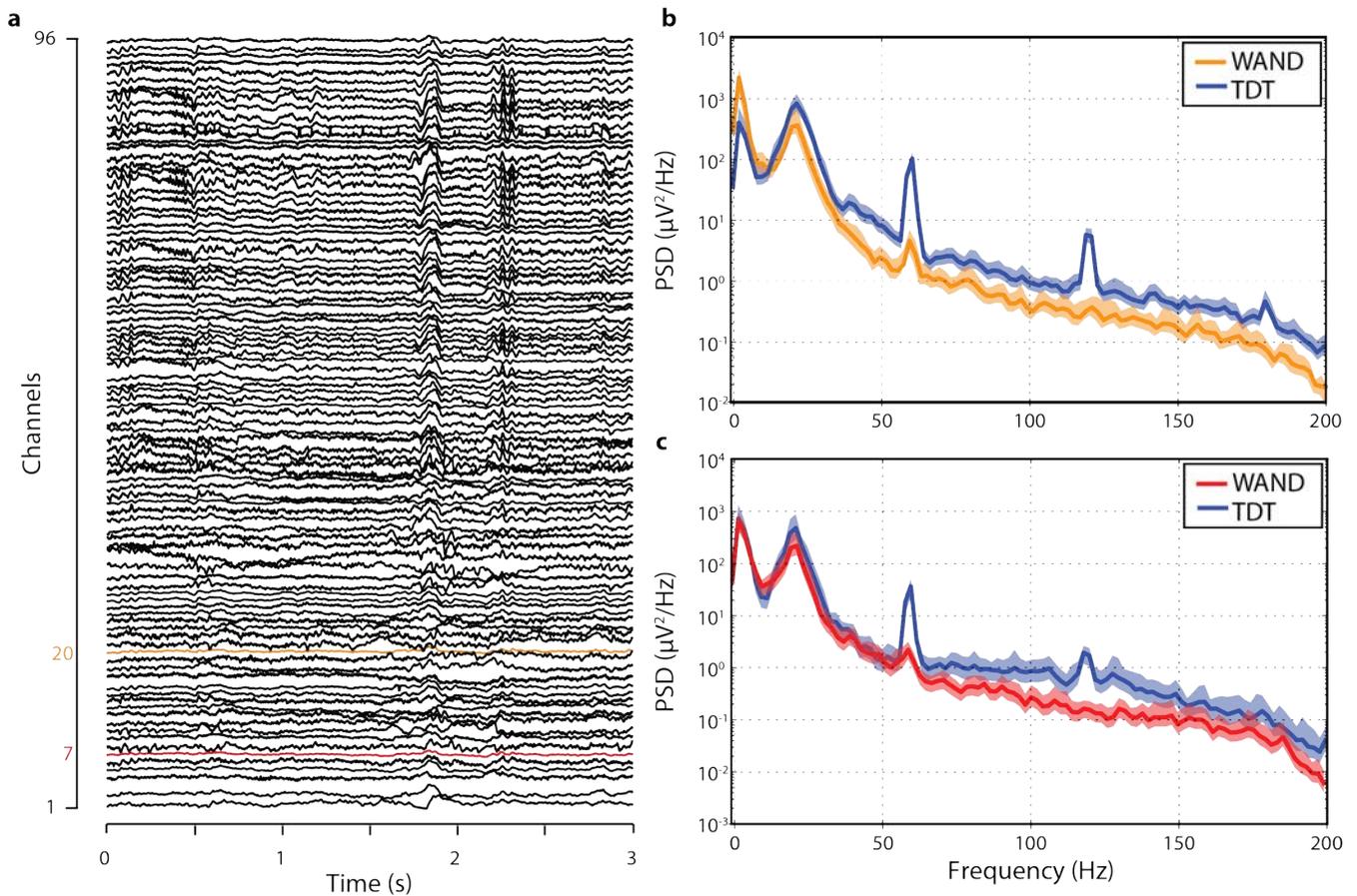

**Figure 2 – Wireless, multi-channel recording.** (**a**) Representative 3-second segments of simultaneous LFP recordings from 96 channels taken during freely moving behavior. (**b, c**) Comparison of power spectral density (PSD) from channel 20 (**b**) and channel 7 (**c**) for recordings taken from WAND and subsequent recordings taken from a commercial wired neurophysiology system (TDT). Error bars are the SD.

To demonstrate robust detection of biomarkers in WAND recordings and establish a baseline for neuromodulation experiments, we recorded LFP activity during a standard self-paced, center-out joystick task (**Fig. 3a, b**). During this behavior, ongoing beta and high-gamma rhythms are inversely modulated by task-related periods of movement (**Fig. 3c, d**). Beta band oscillations are found to emerge during specific motor actions and notably prior to instructed reaches or movements[38,64,65]. In pre-motor and motor areas, this rhythm has been linked to neural activity related to motor preparation[66–69]. The subject had an average reaction time (RT) of 183.3 ± 4.8 (SEM) ms across 400 trials. For LFP signals recorded from pre-motor and motor areas, we found that RT was significantly correlated with the average power of beta band activity around the Go Cue (Pearson's correlation: r = 0.12, p = 0.03).

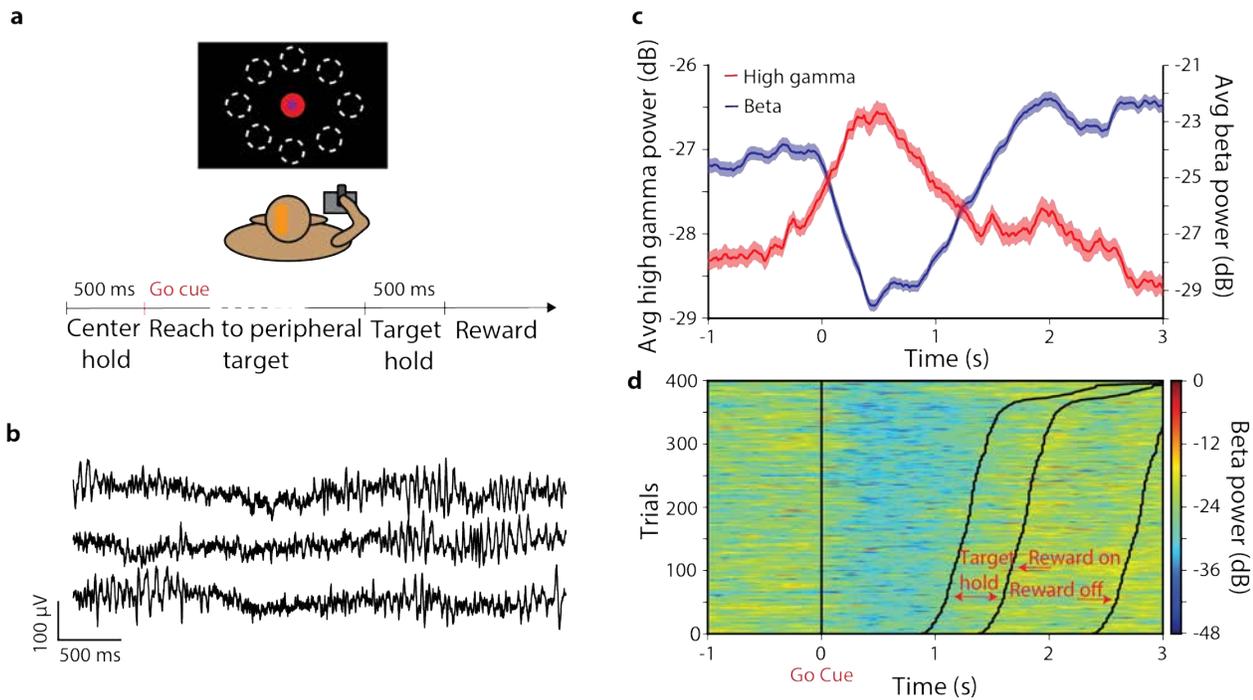

**Figure 3 – LFP recordings during joystick task.** (**a**) Diagram of the center-out joystick task with timeline of task periods for movement and reward. (**b**) Representative LFP recordings from three channels during the center-out task. (**c**) Trial-averaged (n = 400) beta (13 – 22 Hz) and high-gamma (70 – 200 Hz) power aligned to the Go Cue during the center-out task. Error bars are the SEM. (**d**) Beta power aligned to the Go Cue. Each row represents activity from a single trial. Trials are organized by the time to Target Hold following the Go Cue.

To validate long-term, wireless system functionality, we performed unconstrained, overnight recordings for five nights in the subject's home cage, recharging the battery between each session (**Fig. 4a**). We recorded on average 10.2 consecutive hours per night, with a mean packet error rate (PER) below 0.5 % and median PER below 0.1%, where each transmitted wireless packet contained 1 ms of neural data from all streamed channels. Offline analysis of the data revealed useful sleep-related biomarkers. Delta (0 – 4 Hz; **Fig. 4b**) and theta (4 – 7 Hz; **Fig. 4c**) frequency bands are known to have elevated power during sleep states relative to wake states[70,71]. K-complexes are sleep-specific phasic waveforms that occur spontaneously and are observed through the obtained neural recordings during epochs of increased delta power (**Fig. 4d-g**), consistent with classification of sleep state intervals.

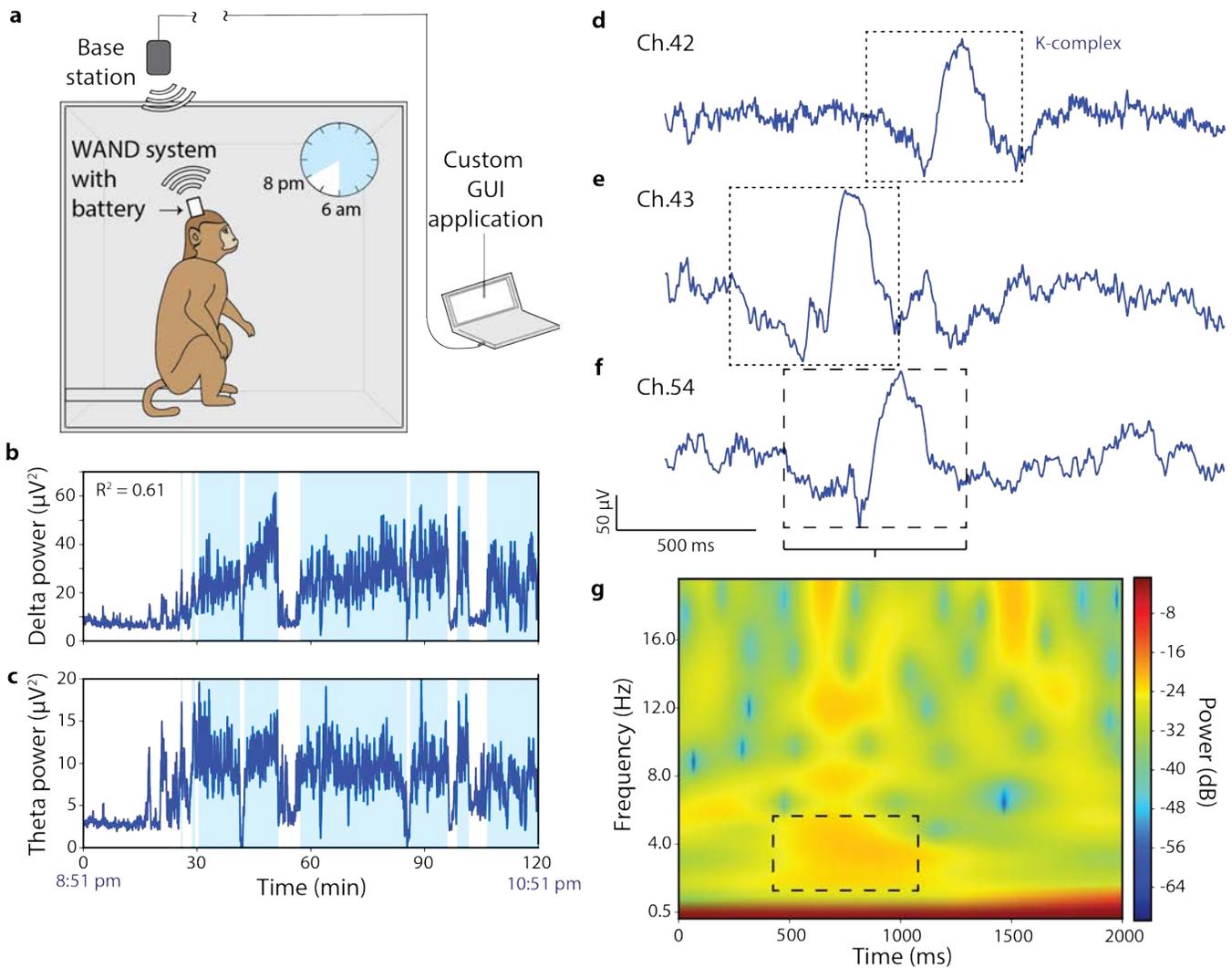

**Figure 4 – Overnight, untethered recording during sleep.** (**a**) Cartoon description of in-cage wireless recordings. (**b, c**) Delta (0.5 – 4 Hz) (**b**) and theta (4 – 7 Hz) (**c**) power from two-hour segment of overnight recording beginning at 8:51 pm. The powers are significantly correlated ($R^2 – 0.61$). K-means was used to classify the activity into states of increased and decreased delta and theta activity, which are indicated by the absence and presence of the light blue background in the plots. (**d, e, f**) Example K-complexes from the caudate (**d, e**) and from the anterior cingulate cortex (ACC) (**f**). (**g**) Spectrogram of activity from channel 54 during the same time window as the waveform shown in (**f**). Increased delta power occurs coincidently with the K-complex.

## Simultaneous recording and stimulation

True simultaneous recording and stimulation is enabled through co-design of the NMIC artifact prevention and mitigation methods with back-end cancellation algorithms running on the on-board microcontroller. To demonstrate WAND's ability to recover neural signals from stimulation artifacts in real-time, we performed experiments delivering open-loop stimulation while recording LFP. For each set of stimulation parameters, we recorded three consecutive segments of LFP: (1) without stimulation, (2) with stimulation turned on but without back-end artifact cancellation, and (3) with artifact cancellation turned on (**Fig. 5a**). During the train of identical bipolar, biphasic stimulation pulses, the segment of LFP with uncancelled artifact (**Fig. 5a**, middle section) demonstrated varying artifact morphology due to the non-integer ratio between the sampling rate and the stimulation frequency (99.8482 Hz shown) (**Fig. 5b**). Stimulation pulses occurring completely within a single sample integration window caused only a single sample direct artifact, while pulses occurring at the boundary between two integration windows caused the direct artifact to last two samples. We calculated averaged templates of single- and double-sample artifacts (**Fig. 5c**) for varying stimulation amplitudes and pulse widths, and confirmed a linear relationship between these parameters and the artifact amplitude (**Fig. 5d**). For all stimulation parameters within our protocol, recorded direct artifacts on the non-stimulating electrodes remained well within the 100 mV linear input range of the front-end amplifiers, despite the high voltages (~10 V) induced on the stimulating electrodes, thus demonstrating saturation-free recording in the presence of stimulation.

Following both single- and double-sample direct artifacts, the indirect artifacts were very small and brief, visible only in the averaged templates (**Fig. 5c** inset). Following a single-sample direct artifact, the indirect artifact was already suppressed to within -60 dB of the peak amplitude by the following sample, and to within the electronic noise floor of 1.6 µVrms by the second sample. Indirect artifacts following double-sample direct artifacts were fully suppressed below the noise floor. These results demonstrate that the recording front-ends rapidly recovered from stimulation pulses, minimizing data distortion.

The residual stimulation artifacts still caused broadband contamination of the recorded spectrum, which we quantified with the ratio, R = 32.78 dB, of signal power integrated form 1 – 200 Hz of LFP during stimulation to baseline LFP (**Fig. 5e, f**). Since the recorded artifacts are short in duration (1-2 samples), we chose to implement a method of linear interpolation for artifact cancellation in the back-end[54]. Samples coinciding with stimulation pulses are flagged by the NMIC, ensuring accurate detection of artifacts. Samples were then buffered in the microcontroller, and artifacts were removed by linearly interpolating between the pre-artifact sample and the sample following the maximum possible direct artifact duration (**Fig. 5b**).

While more sophisticated techniques may be employed, we found that this simple linear interpolation was sufficient to suppress the artifact power below the neural signal spectrum. Interpolation over two samples at 100 Hz in baseline LFP data without stimulation caused no significant degradation of the spectrum (R = 0.0091 dB). Furthermore, this method did not depend on the actual values of the artifacts and was not affected by the varying artifact morphology, which would have complicated and increased convergence times of template subtraction and adaptive filtering techniques. With on-board artifact cancellation enabled, we were able to recover the baseline LFP spectrum for signals recorded during the simulation pulse train with R = -0.60 dB (**Fig. 5a, e, f**).

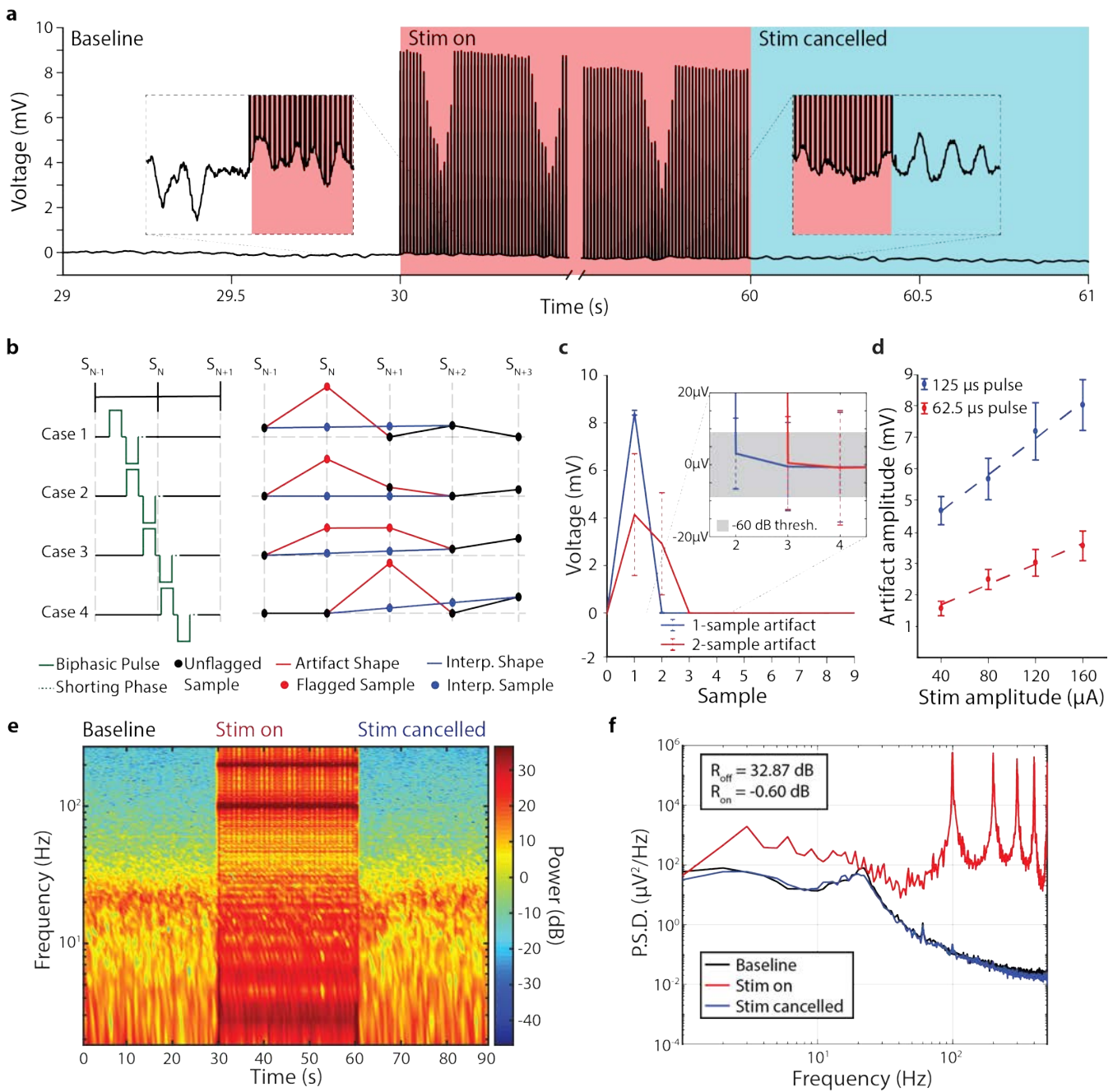

**Figure 5 – Residual artifact analysis and cancellation.** (**a**) 1-second segments of raw signals recorded during different epochs of the open-loop stim experiment: baseline LFP with no stim (white), stim with no artifact cancellation (red), and stim with artifact cancellation (blue). In this example, biphasic pulses were delivered at 100 Hz with 160 µA amplitude and 125 µs pulse width per phase. (**b**) Different cases of relative phase between stim pulse and sampling periods (left) with example resulting samples, artifact flags, and cancelled samples using linear interpolation (right). (**c**) Averaged templates of single- (blue) and double-sample (red) flagged artifacts. The inset is a zoomed portion showing decay of artifact to within -60 dB of the artifact peak (shaded gray). Error bars are the SD (n = 2106 for single-flag and n = 897 for double-flag). (**d**) Average amplitude of artifact from stim amplitudes between 40 µA and 160 µA and pulse widths of 125 µs (blue) and 62.5 µs (red). Error bars are the SD (n = 3003 artifacts). (**e**) Spectrogram of full 90-second recording during the different stim epochs for the example in **Fig. 5a**. (f) Welch power spectral density estimate for each 30-second epoch for the example in **Fig. 5a**.

## *In vivo* biomarker extraction and closed-loop experiment

To further demonstrate WAND's ability to mitigate stimulation artifacts in real-time, and to perform responsive stimulation using on-board computations, we designed a closed-loop stimulation experiment to disrupt movement preparatory activity during a delayed-reach task (**Fig. 6a**). Previous work in macaque monkeys has shown that microstimulation delivered to dorsal premotor (PMd) and primary motor (M1) cortical sites during the delay hold period of a delayed-reach task disrupts preparatory activity and causes an increase in RT[68]. In the study, stimulation was timed synchronous to the task and was not triggered on recorded neural activity. We reproduced this result by detecting periods of preparation (holding) prior to movement using recorded neural activity in M1 and delivering stimulation to electrodes in PMd in response. In this way, stimulation was automatically controlled by WAND, running in a closed-loop manner relying solely on neural activity and completely separate from the task.

Beta band activity (13 – 30 Hz) is known to reflect movement states, with lower beta band power associated with periods of movement and higher beta band power associated with the absence of movement. Thus, we chose beta band power as the WAND control signal for closed-loop classification of hold periods prior to movement. In this way, closed-loop operation of WAND was completely agnostic to the behavior task states, and stimulation delivery relied solely on the control signal. We heuristically selected a policy of delivering a preconfigured stimulation pulse train when both the beta power and it derivative exceeded programmed thresholds during the delayed-reach task (**Fig. 6b**). The pulse train parameters were selected to closely match values used in previous work (333 Hz for 57 ms)[68] within WAND specifications. To avoid stimulation multiple times within the same delay hold period, our policy also incorporated a "dead time" of three calculation periods, or 768 ms. Neural activity is recorded throughout the task, and while stimulation turn-off can be based on neural signature, we chose to adhere to durations used in previous work to demonstrate reproducibility of an established result.

Post-hoc analysis showed that RT increased significantly in behavioral trials when stimulation was delivered during the hold period prior to the Go Cue, relative to trials when it was not (**Fig. 6c**). The increase of 22.0 ms in average RT, consistent with previously reported results for microstimulation delivered in PMd, and the change in the distribution of RT (**Fig. 6d**) indicate that neural preparatory activity was successfully disrupted using our closed-loop neuromodulation approach. This functional change in behavior serves as a representative demonstration of how WAND may be used to compute biomarkers in real-time as part of a closed-loop stimulation paradigm and perform online stimulation artifact mitigation.

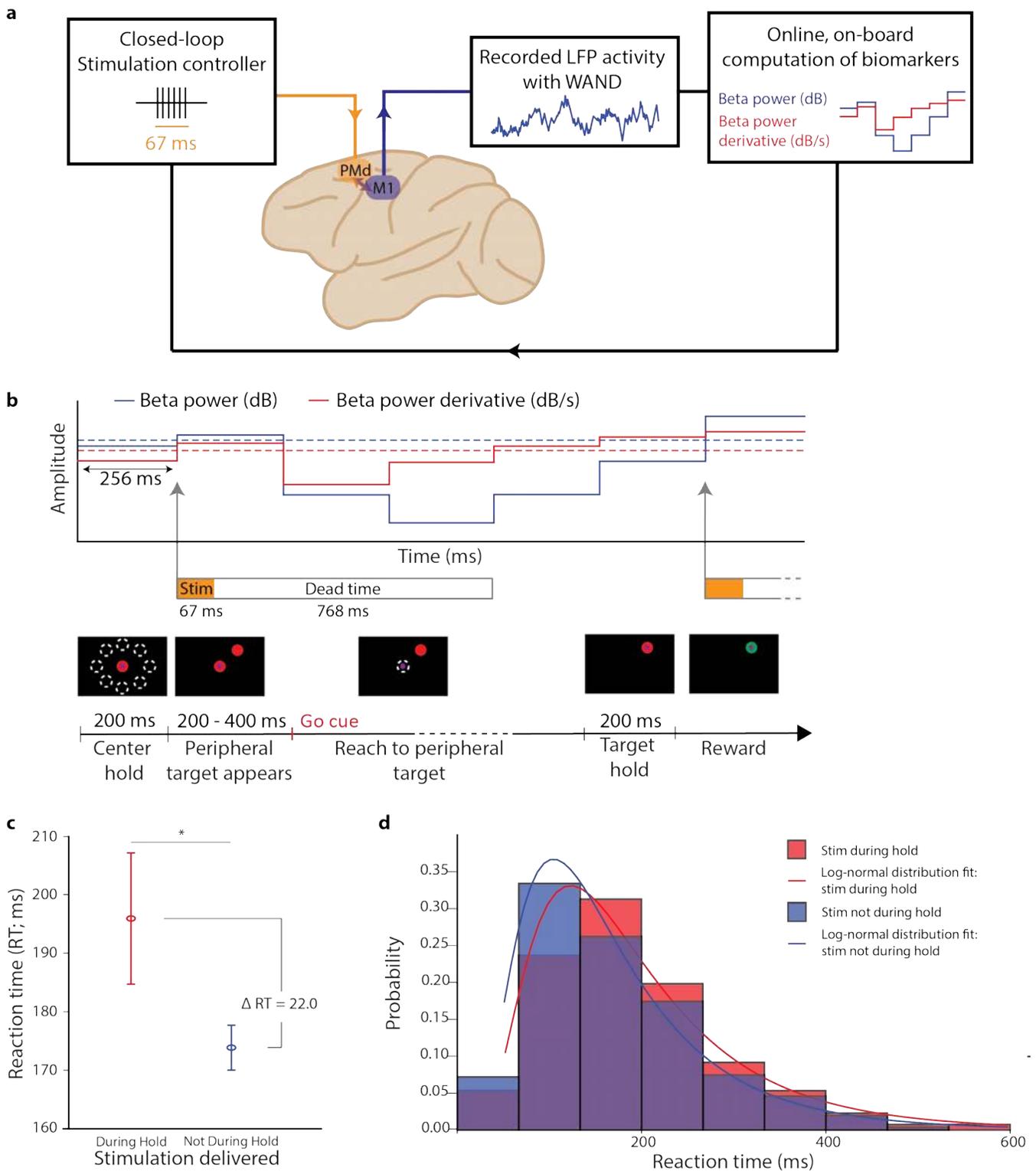

**Figure 6 – *In vivo* closed-loop experiment.** (**a**) Description of closed-loop paradigm, where recorded activity in M1 is used to control stimulation in PMd. (**b**) Diagram of the delayed-reach task and the closed-loop algorithm implemented during this task. Stimulation is delivered when beta power and its derivative exceed their thresholds. (**c**) The trial-averaged RTs for trials (n = 997) in which stimulation was delivered successfully during the hold period compared to when it was not. Error bars are the SEM. Significance was determined using a two-sided Mann-Whitney U-test (U = 44193.5, $n_1$ = 131, $n_2$ = 763, *p < 0.05). (**d**) Normalized RT histograms and log-normal fit to approximate the respective probability density functions.

## Discussion

We have demonstrated WAND, a small form factor, wireless neuromodulation device enabling simultaneous recording and stimulation for a variety of research purposes. Integration of a custom ASIC with an on-board FGPA and radio enables high-quality, long-term multi-channel recording and stimulation during free behavior, full cancellation of stimulation artifacts, flexible programmability, and low-latency processing for delivery of closed-loop microstimulation based on detected biomarkers.

**Table 1** summarizes WAND system specifications and compares them to that of other recently published closed-loop neural interfaces that have been fully packaged and validated *in vivo*. We view WAND as a deployable research tool ready for use with large animals and potentially humans, and we therefore limit our comparisons to similar devices that can operate fully wirelessly and autonomously to make immediate impact on scientific and clinical discovery. Our criteria include fully on-board recording, computation, and stimulation ability; wireless data streaming or on-board memory for data storage; and an implantable or wearable power source. The Neurochip-2[16], PennBMBI[17], NeuroPace RNS[18], and Activa PC + S[15,72,73] devices allow for flexible biomarker detection and triggered stimulation, but with a low number of channels. A device developed at the University of Toronto[19,45,74] enables recording and stimulation on more channels, but all of these devices still stream data at a low rate, preventing large-scale multi-site analysis. WAND improves upon these limitations as the only device to our knowledge that incorporates a large number of recording and stimulation channels, have a wireless data rate to support a large number of streaming channels, and also provide closed-loop neuromodulation capabilities. It is also, to our knowledge, the only neural interface system that actively cancels stimulation artifacts through both hardware and software techniques for completely artifact-free recording during stimulation. Although the Neurochip-2 and Activa PC+S devices utilize some hardware and experimental setup techniques, large residual artifacts still appear to effect performance during stimulation[15,16].

The experiments shown here are intended to outline and demonstrate the capabilities enabled by WAND, paving a path toward use of this technology as a tool in clinical and neuroscientific research. For this work, device form factor, size, channel count, and sensor integration were designed specifically to interface with the microelectrode array implanted in the primate subject. Future research will incorporate other features of WAND, such as the inertial sensor and multi-site stimulation. The architecture of WAND makes it amenable to function as a general-purpose research device, requiring only minor modifications to be re-optimized for new applications. Algorithmic development on the FPGA and microcontroller allows for extracting other neural biomarkers such as band powers or line-lengths used to detect seizure onset in epileptic patients[3], and new closed-loop classification and control algorithms may be conceived for integrating neural activity from a larger number of recording channels. Further research in electrode configuration and improved back-end artifact cancellation algorithms may allow recovery of the full underlying neural signal with no lost samples, a limitation WAND still faces. Currently, the closed-loop control and artifact cancellation algorithms are implemented in the microcontroller, with very little of the FPGA fabric resources (< 5%) being utilized. Preliminary results of porting these algorithms to the fabric indicate that there is still ample room for increasing their complexity to achieve better efficacy.

In a clinical context, the device can be hermetically packaged and used to provide on-demand therapy in deep-brain stimulation while continuously monitoring the neurological response during treatment. For example, in Parkinson's Disease patients treated with DBS, Swann et al.[8] discovered a spectral peak biomarker linked to dyskinesia, an adverse effect of DBS therapy. The closed-loop paradigm we demonstrate in this work can be easily modified for this application, as our current biomarker detection algorithm can already be used to sense this spectral peak. Modifying the control policy to reduce stimulation amplitude rather than trigger stimulation when the peak is detected can thereby reduce the incidence of this undesirable side effect.

**Table 1 – Comparison of closed-loop neuromodulation systems with full *in vivo* validation**

| | Neurochip-2[16] | PennBMBI[17] | U. of Toronto[19,45,74] | NeuroPace RNS[18,75,76] | Activa PC + S[15,72,73,77] | WAND (This Work) |
|---|---|---|---|---|---|---|
| **Dimensions** | 63 x 63 x 30 mm$^3$ | 56 x 36 x 13 mm$^3$ recording and processing; 43 x 27 x 8 mm$^3$ stimulator | 22 x 30 x 15 mm$^3$ | 28 x 60 x 7.7 mm$^3$ | 39 cm$^3$ | 36 x 33 x 15 cm$^3$ |
| **Weight** | 36 (39*) g boards, 145 (204*) g total | -- | 12 g total | 16 g | 67 g total | 7.4 g board, 17.95 g total |
| **Power** | 284-420 mW | 290 mW* | 45 mW | -- | -- | 172 mW |
| **Wireless link** | IR | Nordic Enhanced Shockburst | ZigBee | 20-50 kHz short range inductive | 175 kHz ISM | Nordic BLE |
| **Data rate** | -- | 2 Mbps | 250 kbps | -- | 11.7 kbps | 1.96 Mbps |
| **Real-time streaming** | No | 4 channels | 1 channel* | 1 channel | 2 channels raw, 4 channels compressed | 96 channels + 3 accelerometer channels |
| **# Recording channels** | 3 | 4 | 256 | 4 | 4 | 128 |
| **Recording power/channel** | -- | 1.25 mW/channel | 52 µW/channel | -- | 5 µW/channel | 8 µW/channel |
| **Sampling rate** | 2/24 kS/s (24 kS/s only on 1 channel) | 21 kS/s | 15 kS/s | 250 S/s | 422 S/s | 1 kS/s |
| **ADC resolution** | 8 bits | 12 bits | 8 bits | 10 bits | 10 bits | 15 bits |
| **Artifact mitigation** | Transient gain reduction | None | None | Low pass filter | Front-end filtering, heterodyning, symmetric sensing | DR increase, memoryless sampling |
| **# Stimulation channels** | 3 | 2 | 64 | 8 | 8 | 128 |
| **Max. current** | 5 mA | 1 mA | 250 µA | 11.5 mA | 25.5 mA | 5 mA |
| **Compliance** | ± 15 (50*) V | ± 12 V | 2.6 V | 12 V | ± 10 V | 12 V |
| **Charge balancing** | Matching resistors (0.1 %) | 0.75% mismatch | -- | < 10 µC/sec charge imbalance | Passive discharge | Biphasic current source reuse, 0.016 % mismatch |
| **Artifact cancellation** | No | No | No | No | Stimulation as feature for SVM | Linear interpolation |
| **Biomarker detection** | Spectral power, AP detection | Spectral power, time domain features, AP detection | Phase locking value (PLV) seizure precursor | ECoG signal intensity, line length, half-wave | Spectral power | Spectral power |
| **Closed-loop control** | Detection/threshold triggered | Detection/threshold/sensor triggered | Thresholding | Detection/thresholding | 2D SVM | Thresholding |
| **Animal model** | Primate | Rodent | Rodent | Human | Ovine | Primate |
| ***In vivo* closed-loop paradigm** | Spike triggered | Sensor-node event (button press) triggered | PLV threshold triggered | Seizure detection and responsive treatment | SVM-classification (spectral power, stimulation) triggered | Spectral power threshold triggered |
| **Notes** | * High compliance stimulation version | * Estimated | * Estimated | | * Estimated | |

## Methods

### WAND board components

The WAND board (**Fig. 1b, d**) consists of an SoC FPGA with a 166 MHz ARM Cortex-M3 processor (SmartFusion2 M2S060T, Microsemi) acting as a master module. The FPGA forms a custom 2 Mbps digital signal and clock interface with a pair of NMICs, aggregating data and commands in hardware FIFOs[63]. The Cortex-M3 processor selects which channels are streamed or used for closed-loop, and runs the artifact cancellation and closed-loop algorithms. It connects to a 2 Mbps 2.4 GHz low-energy radio (nRF51822, Nordic Semiconductor) via SPI running at 3.08 MHz to form a bidirectional, half-duplex link with the base station and GUI.

We developed a custom radio protocol using a time division duplex scheme, allowing low-bitrate commands to be sent from the base-station to the board and high-bitrate neural recordings to be continuously streamed out for logging. The exact division between uplink and downlink can be adjusted to suit the application and streaming state, with a maximum effective bitrate of ~1.6 Mbps. Two streaming modes are available. In open-loop mode, 96 channels of data are streamed to the base station. In closed-loop mode, only the control channel and one of the stimulation channels are streamed, along with the calculated power spectral density.

A 20 MHz crystal oscillator provides a clock source to the FPGA and processor, which then generates a 20.48 MHz clock for the NMICs (Cortera Neurotechnologies, Inc.). On-board buck converters (TPS6226x, Texas Instruments) generate the 1.2 V, 1.8 V, 2.5 V, and 3 V supplies needed by the rest of the system from a pair of 4.1 V, 250 mAh Li-ion batteries (ICP521630, Renata). A battery charger IC (LTC4065, Linear Technology) and 3-way connector allows for the battery to be safely charged without disconnecting it from the system.

A 6-axis accelerometer and gyroscope (MPU-6050, InvenSense) and 512Mb low-power SDRAM (MT46H32M16LFBF-5, Micron Technology Inc.) are also connected to the processor through I2C and DDR, respectively, although they are unused in this work.

Device fabrication steps consisted of fabricating the 8-layer PCB, populating board components, wire bonding the NMICs, and soldering the neuro nano-strip connectors (custom order, Omnetics Connector Corp.) for interfacing with the microdrive electrode array. FPGA hardware was written in Verilog, while the Cortex-M3 and radio were programmed in c. A combined JTAG and SWD connector allows users to reprogram and debug SmartFusion2 and radio.

### Base station and software GUI

A wireless base station consisting of a radio (nRF51822 Evaluation Kit, Nordic Semiconductor) and an SPI-to-USB bridge (CP2130EK, Silicon Labs) was used to communicate with WAND (**Fig. 1d**). A custom Python GUI was developed to control and monitor data streamed from WAND on a PC (**Supplementary Fig. 5**). Users can setup the system for multiple use cases, visualize real-time neural recordings, configure all NMIC settings, and configure the closed-loop classification algorithm. Recorded data is saved in HDF5 data format along with relevant use case settings, NMIC configurations, and other notes for the recording.

### Neuromodulation ICs (Cortera Neurotechnologies, Inc.)

The recording subsystem on each NMIC (**Fig. 1c, Supplementary Fig. 3**) is comprised of 64 mixed-signal 15-bit recording channels operating at 1k samples/s. Each recording channel has a selectable input voltage range of 100mV or 400mV allowing simultaneous amplification and digitization of the electrode offset, neural signal, and stimulation artifact within the linear range.

The four on-chip stimulators can be multiplexed to any of the electrodes and allow for a variety of programmable stimulation parameters, including current amplitudes, pulse timing, and frequencies (**Supplementary Fig. 1**). Stimulation pulses are delivered in 3 phases: a setup phase with configurable setup time, a pulse phase configurable to be mono-phasic or biphasic with configurable pulse widths and interphase gap, and a shorting phase with a configurable shorting time where electrodes are shorted to the reference. The NMIC assists in artifact cancellation by flagging all samples coinciding with any of the stimulators being active; however, samples coinciding with the shorting phase only are not flagged and care must be taken to also remove artifact from those samples (**Fig. 5b**). The artifact flag is implemented as a single bit appended after the most significant bit of the 15-bit ADC value, creating a 16-bit value per sample per channel. To enable low-latency (sub-ms), highly-programmable stimulation (225 bits of customization), the NMIC uses double content shadow registers, meaning stimulation parameters can be changed while the previous stimulation pulse or waveform is executed. A low-overhead command initiates a programmed stimulation pattern.

On-chip programmable DC-DC converters provide a 1V supply to the recording and digital circuits as well as a selectable 3/6/9/12V supply to the stimulator, adjusting the compliance for different stimulation regimes for improved power efficiency (**Supplemental Fig. 3**). All power management is therefore integrated on the chip, enabling power from a single supply without the need for large off-chip power conversion circuits[27].

**Artifact cancellation and open-loop experiment analysis**

Frames of concurrent 16-bit samples from the enabled NMIC recording channels arrive at the FPGA every 1 ms. Because some unflagged samples may still be affected by the shorting phase, our artifact cancellation always interpolates over the maximum number of consecutive flagged samples possible after detection of the first artifact sample (**Fig. 5b**). This can be calculated by finding the length in milliseconds of the entire pulse, rounding it up to the nearest integer, and adding 1. For biphasic 125 μs pulses with a 31.25 μs interphase gap and 31.25 μs shorting phase, the length of the pulse is 0.3125 ms, which requires 2 samples to be cancelled per artifact.

Artifact cancellation is implemented on the ARM Cortex-M3 processor before packetization of data for wireless transmission or use in the closed-loop algorithm. Eight frames of samples are buffered, allowing cancellation of artifacts lasting up to seven frames. Artifacts are detected upon finding the first frame with a set artifact flag and cancelled once the first clean frame is received. Because of the 8-frame buffering, there is a delay of 8 ms between frames being received by the FPGA and frames being transmitted to the base-station or being used for closed-loop.

**Closed-loop algorithm**

The closed-loop control algorithm is implemented in the ARM Cortex-M3 processor and triggers stimulation based on real-time spectral analysis of any one of the 128 recording channels (**Fig. 6a, b**). We compute the power spectrum of buffered windows of data using the fixed-point FFT and magnitude squared functions of the ARM CMSIS DSP library. Each window is demeaned and scaled 64x before computation. The window length, N, can be configured from the GUI to be any power of 2 between 16 and 2048, and successive windows overlap by N/2 samples.

From the power spectrums, we can derive up to two control signals. Each control signal can either be the integrated power across a specified frequency band, or the derivative of that power estimated by subtracting the newly calculated power value from the previous one. For each control signal we can specify the threshold for either the power or derivative. After each calculation, the decision to stim can either be the logical AND or logical OR of the threshold crossings from each control signal. A programmable "dead time" can be applied to prevent stimulation being triggered by consecutive power measurements. An additional random control mode triggers stimulation pulse trains at pseudorandom intervals between configurable minimum and maximum time intervals.

**Surgery and electrophysiology**

A customized semichronic microelectrode array (Gray Matter Research, Inc.; Bozeman, MT) was implanted unilaterally in one male rhesus macaque (weight ~9.1 kg) (**Fig. 1a, d**). The subject was implanted unilaterally with the custom-machined chamber enabling access to pre-motor and motor cortical regions. The chamber position was calculated based on images obtained from 1.5-T magnetic resonance imaging (MRI) scans of the subject's brain. The semichronic array features a titanium chamber form-fitted to the cranium of the subject and a microdrive housing 157 single microelectrodes that are independently moveable in the depth axis. The microdrive sits within the implanted chamber and a sterile seal for the system is maintained. The microelectrodes are gradually lowered into neural tissue over time and their positions are adjusted throughout the experiment to better isolate neural activity in the nuclei of interest. Electrode positions are controlled by miniature screw driven actuators traveling along threaded rods. Electrical contact with the electrodes is achieved through a printed circuit board (PCB) and Omnetics headers are used to connect the PCB to neural recording systems, such as WAND or standard tethered electrophysiology equipment. There were two types of microelectrodes used in the semichronic array. The first were tungsten electrodes with epoxylite insulation (500 – 800 kOhm; FHC), a standard electrode type for acute neural recording experiments. The second type were platinum-iridium (PtIr) electrodes with parylene-C insulation, which are standard for neuromodulation experiments (200 – 350 kOhm; Microprobes; Alpha Omega). Electrical stimulation in this study was exclusively performed using the PtIr microelectrodes. All experiments were performed in compliance with the NIH Guide for the Care and Use of Laboratory Animals and were approved by the University of California, Berkeley Institutional Animal Care and Use Committee (protocol AUP-2014-09-6720).

**Primate experimental procedures**

Overnight recordings were carried out with the subject moving freely throughout the home environment and were typically taken from approximately 8 pm to 6 am (**Fig. 4a**). The subject was pair-housed with an NHP cagemate and was in social contact with the cagemate throughout the recording session. It worth noting that the dimensions of the homecage environment could accommodate up to four NHPs, and thus it is feasible to utilize WAND for recording from a small population of socially housed animals without compromising the streaming wireless signal integrity. The base station receiver was mounted on the ceiling approximately 0.5 m from the top of the cage and was connected to a computer running the custom GUI application for acquiring the neural recordings.

The subject was also trained in a standard center-out joystick task and a delayed-reach joystick task for in-chair behavioral recordings and for the closed-loop experiment (**Fig. 3a, 6b**). Both tasks were self-paced. Briefly, the subject was trained to use a joystick to control a cursor on a computer screen and move to circular targets presented on the screen. The joystick was affixed to the front of the primate chair and the subject was free to use either hand at any point in the task to control the joystick.

In the center-out task (**Fig. 3a**), a trial begins with the subject holding the cursor at a center circular target for 500 ms. Following this hold period, a peripheral target appears at one of eight target locations equally distributed around the center target at a distance of 10 cm and the center target is removed from the screen, acting as a "go cue". The subject then moves the cursor (i.e. "reaches") to the peripheral target and holds at this target for another 500 ms. If successful, the subject is administered a small juice reward lasting 800 – 1000 ms. A trial was considered successful if the subject completed the two hold periods within a 10 s period.

The sequence of events in the delayed-reach joystick task (**Fig. 6b**) is similar to the center-out task, with the exception being that the peripheral target appears prior to the "go cue", which is signaled with the disappearance of the center target. The hold period for the center target lasts 400 ms before the peripheral target is shown. This initiates the "delay period" with a duration that varied randomly trial-by-trial with a range of 200 – 400 ms. After the delay period, the center target disappears from the screen signaling the "go cue" and the subject is cued to reach to the peripheral target. The range of delay durations was chosen to allow for movement preparation and to ensure that microstimulation occurred near the go cue for a nontrivial number of trials.

**Open-loop artifact cancellation experiment**

The open-loop artifact cancellation experiments consisted of continuous recordings made with 30 seconds of no stimulation, 30 seconds of stimulation with no artifact cancellation, and 30 seconds of stimulation with artifact cancellation for each set of stimulation parameters. Biphasic stimulation, with amplitudes swept in 40 µA steps between 40 µA and 160 µA, were delivered for pulse widths of 125 µs and 62.5 µs and with 100 Hz and 20 Hz stimulation frequencies. Stimulation electrodes were chosen to be the same ones for the closed-loop experiment. During the open-loop stimulation experiments, the monkey was in-chair and did not perform any tasks.

Artifacts recorded with back-end cancellation disabled were sorted offline into 10-sample windows aligned with the sample 0 being the clean sample before artifact starts and sample 1 being the first flagged sample of the artifact (**Fig. 5c**). These segments were used to analyze the size and consistency of recorded artifacts. Offsets were then subtracted from each window such that sample 0 was 0 V. Artifact amplitude was calculated as the average sum of the magnitudes of samples 1 and 2. Artifact duration was then calculated as the average number of samples for which the magnitude was greater than -60 dB of the maximum calculated artifact amplitude.

To determine the effectiveness of cancellation, the power spectrum for each epoch of stimulation and artifact cancellation was estimated using Welch's averaged modified periodogram method with 1000 Hanning windowed samples and overlaps of 500 samples.

**Closed-loop experiment**

For the closed-loop experiment, we used a window length of N = 512 to calculate beta power (13 – 30 Hz) and the derivative of beta power as our control signals. Stimulation was enabled when both the beta power exceeded 33 µV-rms and the delta of the beta power exceeded 10.45 µV-rms. The dead time was set to 3 power calculation windows, or 768 ms (**Fig. 6b**). Biphasic, bipolar stimulation was delivered to PtIr electrodes 52 and 53 in pMD. Stimulation pulses were 160 µA in amplitude with 125 µs pulse widths and 31.25 µs shorting phase. Pulse trains were 18 pulses long and delivered at 256 Hz.

Reaction time was defined as the length of time following the Go Cue for the cursor speed to first achieve a threshold of 5 cm/s. RTs below 50 ms were discarded as they were likely due to the subject initiating movement prior to the Go Cue and RTs above 1 s were also not considered as they indicate a low level of engagement in the task.

**Data and code availability**

The data and code that support the findings of this study are available from the corresponding author upon reasonable request.

**Acknowledgments**

This work was supported in part by the Defense Advanced Research Projects Agency (W911NF-14- 2-0043) to R.M., J.M.R., and J.M.C. and the National Science Foundation Graduate Research Fellowship Program (Grant No. 1106400) to A.Z.  The authors thank Prof. Elad Alon, Dr. Simone Gambini and Dr. Igor Izyumin for technical discussion.

**Author contributions**

A.Z., B.C.J., G.A., A.M., F.L.B., J.M.R., and R.M. designed and tested the system. B.C.J. and R.M. designed and tested the integrated circuits. S.R.S. and J.M.C. designed the *in vivo* experiments. A.Z., S.R.S., B.C.J., G.A., and A.M. performed the experiments and analysis. J.M.R., J.M.C., and R.M. oversaw the project. A.Z., S.R.S., B.C.J., G.A., A.M., J.M.R., J.M.C., and R.M. wrote and edited the paper.

**Competing financial interests**

Authors B.C.J, J.M.R., J.M.C, and R.M. have financial interest Cortera Neurotechnologies, Inc. Cortera Neurotechnologies, Inc. has filed a patent application on the integrated circuit used in this work.

# WAND: A 128-channel, closed-loop, wireless artifact-free neuromodulation device

## Supplementary information

### NMIC stimulator programmability

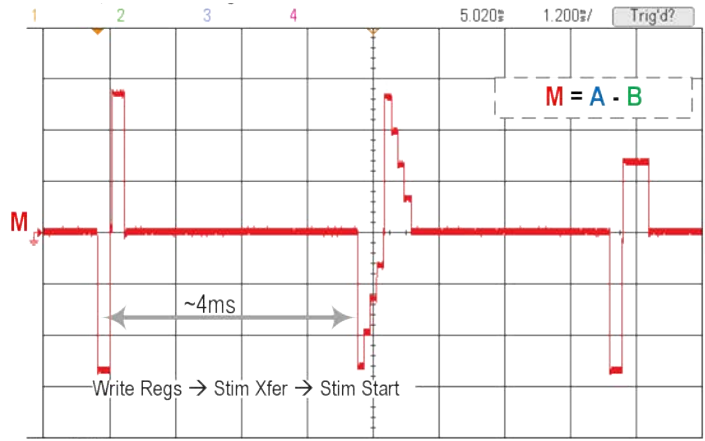

| Parameter | Value | Unit |
|---|---|---|
| **Stimulation subsystem** | | |
| Nominal supply voltage | 3, 6, 9, or 12 | V |
| Stimulation compliance | 11.8 | V |
| Number of stimulation units | 4 | |
| Number of addressable channels | 66 | |
| **Stimulation settings** | | |
| Stimulation current resolution | 20, 40, 60, or 80 | µA |
| Maximum stimulation current | 5.04 | mA |
| Pulse resolution | 15.625 | µs |
| Maximum pulse width | 500 | µs |
| Interphase gap | 31.25-1000 | µs |
| Short time | 31.25-1000 | µs |
| Frequency | 15-255 | Hz |

**Supplementary Figure 1 – NMIC stimulator parameters and rapid waveform update.** A table including stimulator subsystem specifications and settable stimulation pulse parameters is shown on the left. Rapid waveform reconfiguration between pulses delivered at the maximum frequency (255 hz) is demonstrated on the right. Stimulation parameters can be updated every pulse. Additionally, multiple current sources can be multiplexed to a single electrode to create higher amplitude stimulation or complex waveforms, for example.

### System artifact resiliency

Methods implemented on the NMIC for preventing and mitigating stimulation artifact are described here. While the NMIC itself does not completely remove stimulation artifacts, it avoids saturation and minimizes artifact duration to make back-end cancellation more manageable.

In order to prevent large, persistent artifacts, biphasic stimulation pulses should be used instead of monophasic pulses. Following a single-phase stimulation pulse, a second phase pulse with opposite polarity is delivered to actively clear the charge delivered during the first pulse. The amount of indirect artifact is related to the amount of residual charge left over after both phases. The NMIC stimulator architecture prevents large indirect artifacts from occurring by ensuring accurate charge balance between the biphasic pulses and also enabling passive discharge through shorting of electrodes. Common methods for charge-balancing stimulation pulses include current copying[78,79], current driver feedback control[26], and voltage offset correction through feedback[80]. The NMIC employs an H-bridge architecture allowing the reuse of the same current source during both phases of stimulation, eliminating the effects of PVT variations and achieving 0.016% mismatch between phases (**Supplementary Fig. 2**)[27,45,81]. Further details can be found in the NMIC paper[27].

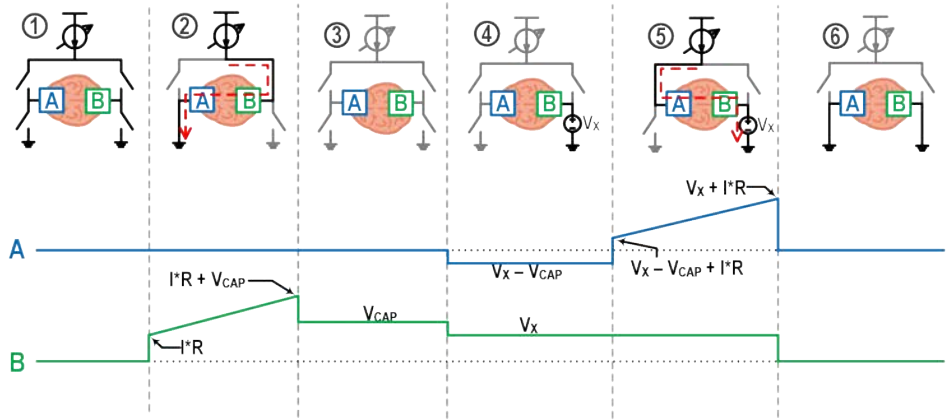

**Supplementary Figure 2 – H-bridge stimulator timing diagram.** The stimulator reuses the current source for both stimulation phases to achieve precise phase matching. The voltage offset, $V_x$, is used during the second phase to prevent the voltage on either electrode from going negative. A shorting phase of programmable duration is used to clear any residual charge on the electrodes.

Because the NMIC integrates the stimulators and front-ends onto the same chip, a common ground reference is shared between these two subsystems (**Supplementary Fig. 3**). This provides an additional benefit that, during passive discharge following a stimulation pulse, the stimulation electrodes are shorted to the reference voltage of the front-ends.

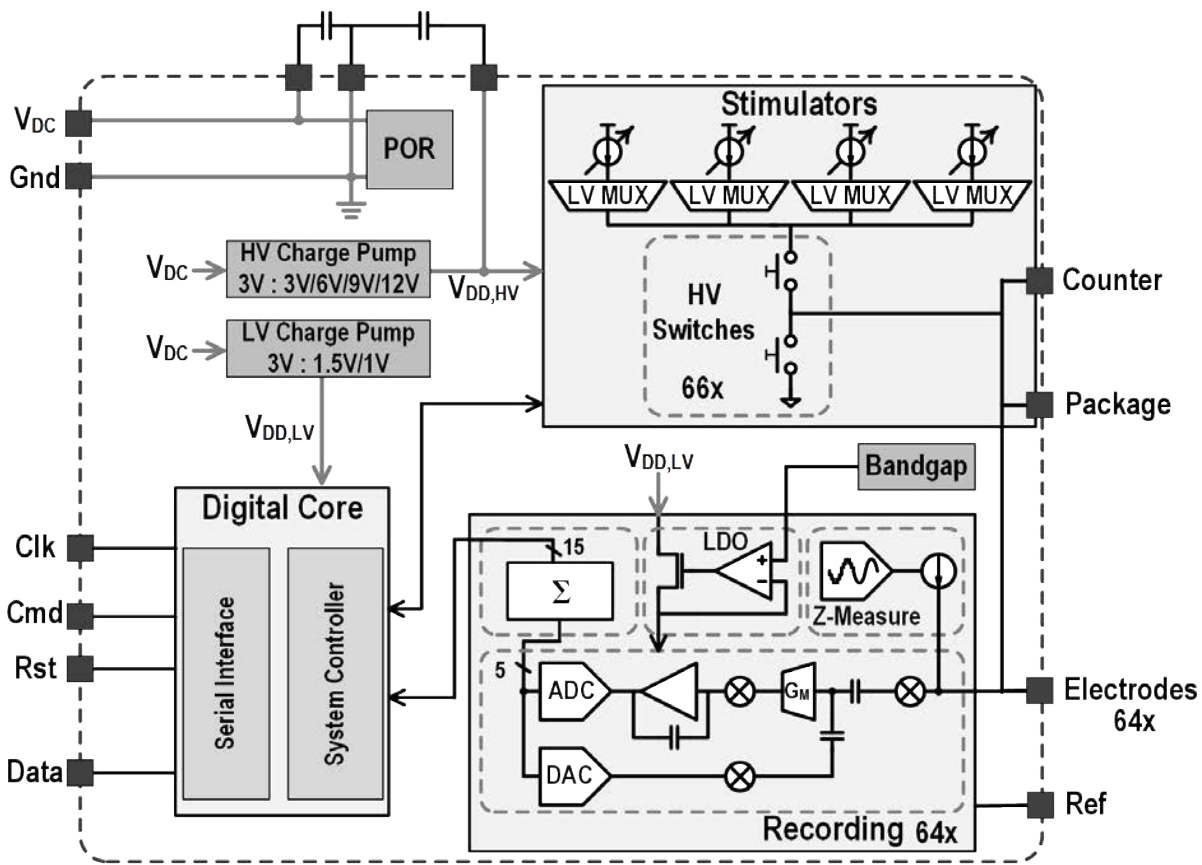

**Supplementary Figure 3 – system architecture of the NMIC.** The NMIC enables a bidirectional neural electrode interface with integrated stimulation, recording, impedance measurement, and power management. The NMIC uses a reference electrode as a common reference for sensing. The counter and package electrodes can be used as additional simulation channels and are unused in this work.

Large direct artifacts and remaining indirect artifacts can saturate front-end amplifiers with small input ranges. Techniques for mitigation of stimulation artifacts at the recording front-end fall into two categories: saturation prevention and rapid recovery[82]. Saturation prevention can be achieved by increasing dynamic range or by subtracting artifact signals in the analog domain. Subtraction of learned artifacts can reduce the voltage swings at the amplifier input, but current implementations still allow residual artifacts to pass through due to error in the subtracted artifact templates[48,49]. Alternatively, designers can increase the amplifier input ranges to tolerate larger signal swings without subtraction[22,27,47]. Rapid recovery methods reduce artifacts due to charge build-up of components in the front-end circuit, with resetting capacitive elements in the front-end being a common method[26,27,50,51]. A more complex implementation involves active discharging of these elements[83]. The NMIC implements a front-end architecture that provides both a large input range and rapid recovery between samples[27].

The NMIC front-end uses a first order, continuous-time, incremental (resetting) ADC with capacitive feedback to achieve a large input range (**Supplementary Fig. 4**). The forward signal path consists of an active integrator and a 5-bit SAR ADC. By bringing the ADC inside the amplifier's capacitive feedback loop, internal signal swings are reduced, resulting in improved linearity. The analog forward path filters the SAR residue providing first order quantization noise shaping, which is approximately 6 dB (1 bit of resolution) per doubling of oversampling frequency. The SAR is oversampled 1024 times resulting in a nominal resolution of 15 bits at 1 kS/s. A reset occurs every output sample using the switches labeled $S_{RST}$. Every sample is memoryless and therefore rapidly recovers from a saturating signal.

The input range is defined by the closed-loop capacitive ratio and the voltage supply of the feedback DAC (930 mV). On each side of the front-end, the input capacitance is 1.6 pF and the effective feedback capacitance is 86 fF, or approximately 2.8 fF per bit of feedback from the DAC. A capacitive attenuation network is used to achieve a feedback capacitance smaller than unit size (31 fF) to improve matching. The input range is calculated as 930 mV · 86 fF / 1.6 pF = 50 mV for each polarity, which is 100 mVpp. A setting to change the capacitance ratio results in an input range of 400 mVpp, which can be used in rare cases when artifact amplitudes exceed 50 mV amplitudes.

This large, linear input range allows the NMIC to record artifacts 10's of mV in amplitude without saturating, while simultaneously recording small neural signals with low noise. The noise level of the front-end is dominated by the thermal noise of the analog forward path and the quantization noise of the converter (nominally about 950 nV$_{rms}$ each, input-referred). Flicker noise of the analog forward path would dominate by an order of magnitude; however, the forward path uses a chopper to up-modulate the flicker noise of the transconductance stage, which is filtered by the integrator.

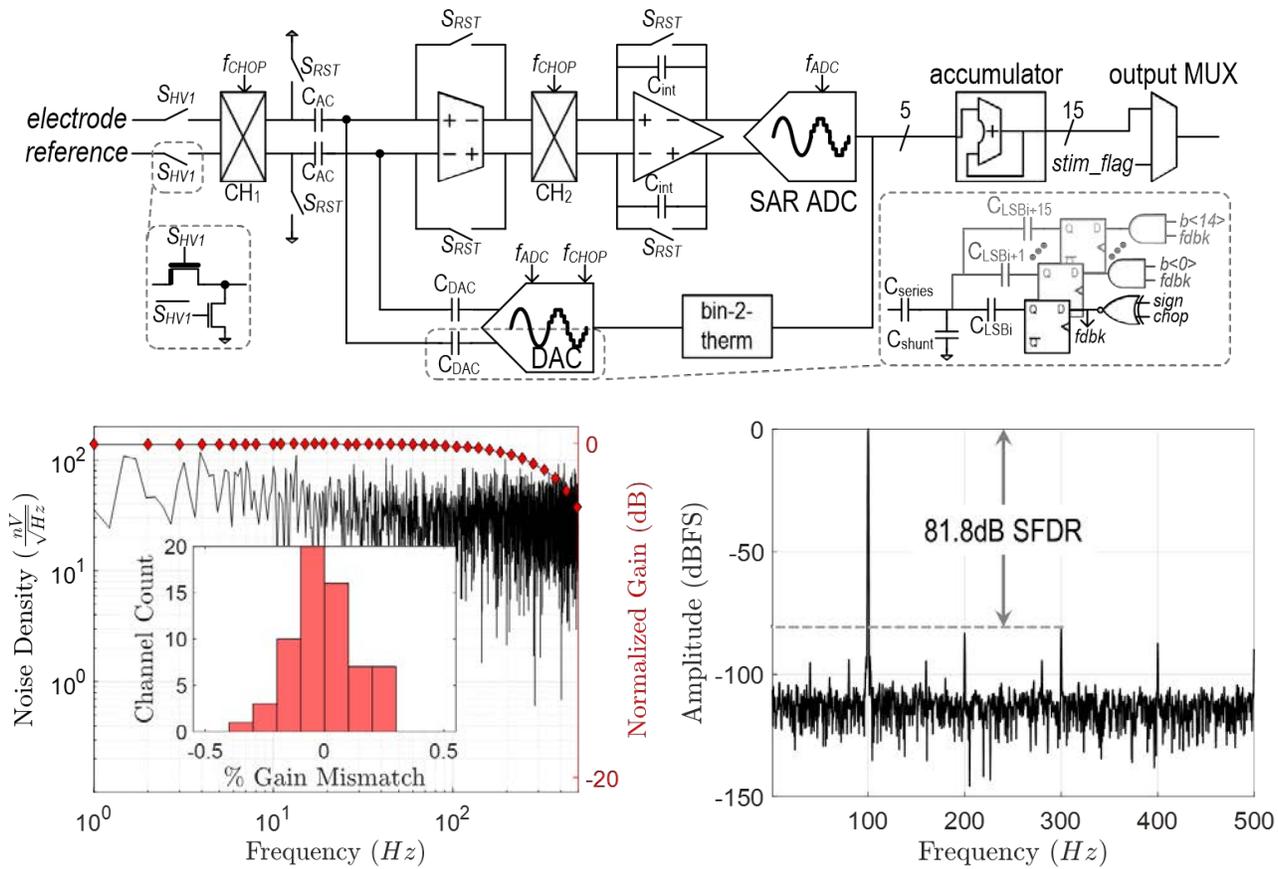

**Supplementary Figure 4 – architecture of the recording front-end and measurement results.** The frontend is an incremental oversampled converter with 15-bit nominal resolution and is DC-coupled. The front-end resets every sample (1ms) which aids in rapid recovery from stimulation artifact. Noise spectrum is input-referred and SFDR spectrum from a single channel is normalized to full scale input (100mV$_{pp}$, f$_{in}$ = 100Hz, SNDR = 76.7dB). All system circuits were active during measurements and the outputs were taken from digital bit stream.

A comparison of fully integrated, bidirectional neural interfaces with an emphasis on artifact prevention and mitigation is given in **Supplementary Table 1**.

# Supplementary Table 1 – A comparison of bidirectional neural interface ICs

| | Rhew JSSC '14[25] | Chen JSSC '14[26] | Shulyzki TBioCAS '15[45] | Mendrela JSSC '16[48] | Kassiri JSSC '17[22] | Johnson VLSI '17[27] |
|---|---|---|---|---|---|---|
| **Process technology (nm)** | 180 | 180 | 350 | 180 | 130 | 180HV |
| **No. channels (rec/stim)** | 4/8 | 8/1 | 256/64 | 8/4 | 64/64 | 64 / 4** |
| **ADC resolution** | 8 | 10 | 8 | 10 | -- | 15 |
| **ENOB** | 5.6 | 9.57 | 5.1 | -- | 11.7 | 12.45 |
| **Sampling rate (samples/s/ch)** | 25k | 62.5k | 15k | 4k | 1k* | 1k |
| **Power/channel (µW)** | 61.25 | 58 | 52 | 0.33 | 0.63 | 8 |
| **IR noise (nV/rtHz)** | 81 | 63 | 113 | 68 | 51 | 68§ |
| **CMRR (dB)** | -- | -- | 60 | -- | 89 | 85 |
| **Input range (mVpp)** | 1.2 | 10 | -- | 10* | -- | 100 / 400 |
| **THD** | 0.8% (1.2mVpp)* | -- | 0.8% (1mVpp) | -- | -- | 0.012% (100mVpp) |
| **Stim max amplitude** | 4.2 mA / 5 V | 30 µA / 10 V | 250 µA / 2.6 V | -- | 1.35 mA / 3.1 V | 5.04 mA** / 12 V |
| **Input coupling** | AC-coupled | AC-coupled | AC-coupled | AC-coupled | Rail-to-rail DC | ±50mV / ±200mV DC |
| **Biomarker Computation** | Logarithmic DSP, LFP energy PI controller | FFT, ApEn, LLS classifier DSP | Off-chip‡ | -- | Phase-synchrony DSP | Off-chip† |
| **Wireless data streaming** | 2.4 GHz backscatter | MedRadio band OOK | Off-chip‡ | -- | UWB | Off-chip† |
| **Power supply** | 915 MHz RF harvesting + 5V battery | 13.56 MHz inductive link | Battery‡ | -- | 1.5 MHz inductive link | Battery† |
| **Artifact duration (ms)** | -- | -- | -- | 2 (assumed) | -- | 1 |
| **Artifact prevention** | -- | Current driver feedback controller | Current driver reuse for balancing | -- | Passive discharge, stimulator matching | Compatibility, current source reuse for balancing, passive discharge |
| **Artifact mitigation** | -- | Saturation detection and rapid recovery | -- | Adaptive filter artifact subtraction, common averaging reference | Rail-to-rail DC offset | High dynamic range, rapid recovery |

\* Estimated
\*\* All four stimulators can be combined on a single channel for 20.16mA
§ Average noise across channels is 77nV/rtHz
† System described in this work
‡ System described in main comparison table

# Graphical user interface

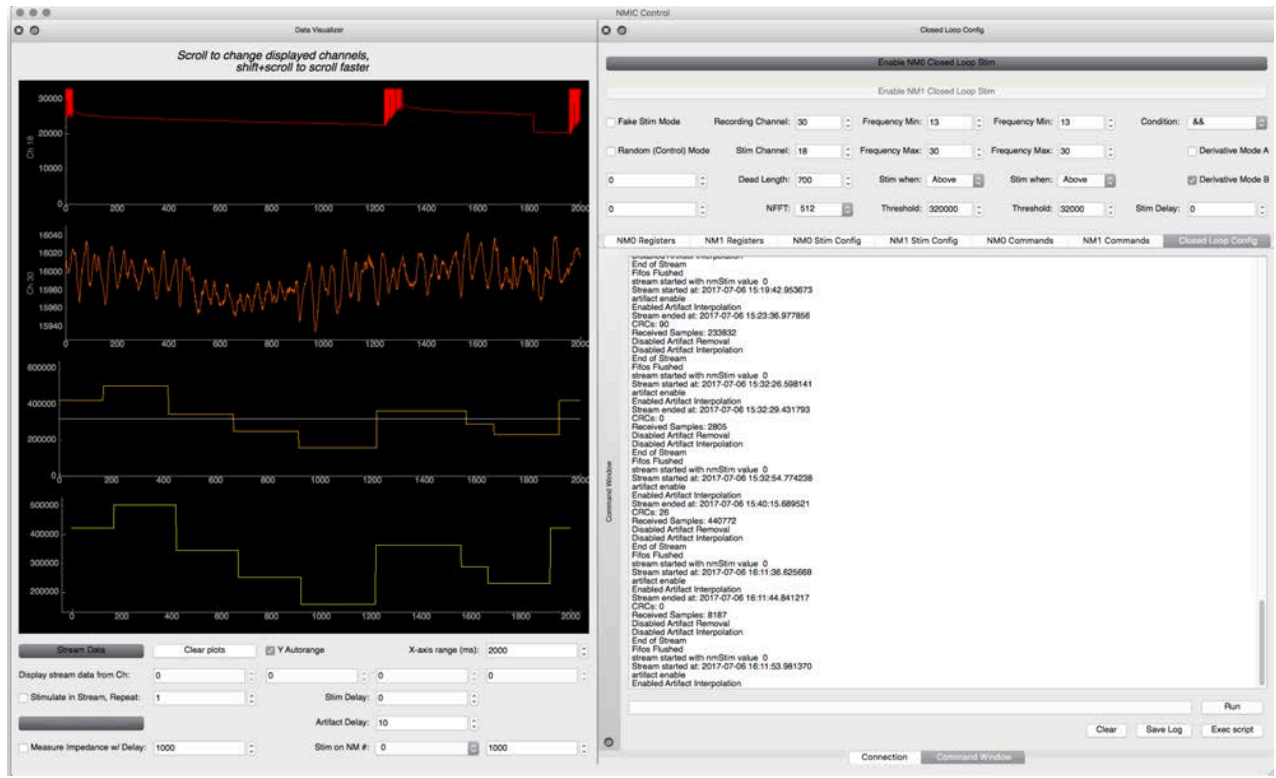

**Supplementary Figure 5 – screenshot of WAND GUI running a closed-loop experiment.** The left side of the window is dedicated to real-time data visualization. From the top, the traces are of one of the stimulation channels, the control channel, and two calculated biomarkers with thresholds shown as a horizontal line. In this case, the two bottom traces appear to show the same information because the both the raw value and the time derivative of the same biomarker were used. Full power spectrums were calculated on the device and transmitted to the GUI, where local integration of band power was performed before visualization. The right side of the window contains settings for the closed-loop experiment and a command shell for displaying data stream information.